\documentclass[aps,prl,twocolumn,superscriptaddress,nofootinbib]{revtex4-1}
\usepackage[pdftex]{graphicx} % arxiv
\usepackage{color} % arxiv

\usepackage[pdftex,bookmarks=true,bookmarksnumbered=true,colorlinks=true,linkcolor=blue,citecolor=blue,filecolor=blue,urlcolor=blue]{hyperref}  % arxiv

\usepackage{amsfonts}
\usepackage{amsmath}
\usepackage{bm}
\usepackage{mathrsfs}
\usepackage{here}
\usepackage[normalem]{ulem}
\newif\ifoneauthor
\oneauthortrue
\newcommand{\unit}[1]{\ {\rm #1}}

\newcommand{\Eq}[1]{Eq. (\ref{#1})}

\newcommand{\Fig}[1]{Figure \ref{#1}}
\newcommand{\HST}{\mathcal{H}_{\text{ST}}}

\newcommand{\tc}{t_{\mathrm{c}}}
\newcommand{\phic}{\phi_{\mathrm{c}}}
\newcommand{\dIs}{\{d_I\}^N_{I=1}}

\definecolor{mygreen}{RGB}{0,115,0}

\DeclareMathAlphabet{\mathpzc}{OT1}{pzc}{m}{it}
\usepackage{here} %図を強制的にその位置に出力するためのpackage \begin{figure}[H]でできる。
\definecolor{gray}{gray}{0.4}

\begin{document}

%\preprint{}
%\twocolumn
%\title{Test of gravitational wave polarizations with inspiral waveforms \\of compact binary coalescences}
\title{Scalar-tensor mixed polarization search of gravitational waves}



\author{Hiroki Takeda}
\email[]{takeda@tap.scphys.kyoto-u.ac.jp}
\affiliation{Department of Physics, Kyoto University, Kyoto 606-8502, Japan}
\affiliation{Department of Physics, University of Tokyo, Bunkyo, Tokyo 113-0033, Japan}
\author{Soichiro Morisaki}
\affiliation{Department of Physics, University of Wisconsin-Milwaukee, Milwaukee, WI 53201, USA}
\author{Atsushi Nishizawa}
\affiliation{Research Center for the Early Universe (RESCEU), School of Science, University of Tokyo, Bunkyo, Tokyo 113-0033, Japan}

%\thanks{}

%\author{David J. Ottaway}
%\email[]{david.ottaway@adelaide.edu.au}
%\thanks{}
%\affiliation{ Department of Physics and The Institute of Photonics and Advanced
%Sensing, The University of Adelaide, Adelaide, South Australia, Australia}


\date{\today}

\begin{abstract}
 An additional scalar degree of freedom for a gravitational wave is often predicted in theories of gravity beyond general relativity and can be used for a model-agnostic test of gravity. In this letter, we report the direct search for the scalar-tensor mixed polarization modes of gravitational waves from compact binaries in a strong regime of gravity by analyzing the data of GW170814 and GW170817, which are the merger events of binary black holes and binary neutron stars, respectively. Consequently, we obtain the constraints on the ratio of scalar-mode amplitude to tensor-mode amplitude: $\lesssim 0.20$ for GW170814 and $\lesssim 0.068$ for GW170817, which are the tightest constraints on the scalar amplitude in a strong regime of gravity before merger.
\end{abstract}

%\keywords{}

\maketitle


\section{Introduction}
General relativity (GR) has passed all experimental and observational tests so far \cite{Will1993, Stairs2003, Murata2015}. However, many alternative theories of gravity have been suggested and studied for various theoretical motivations \cite{Will2005, Clifton2012}. For example, a scalar-tensor theory \cite{Brans1961, Fujii2003} is one of the simplest extensions of GR and some models are motivated from the late-time accelerating expansion of the Universe \cite{Perrotta1999}. It is well known that a gravitational wave (GW) can have two tensor degrees of freedom in GR,
%%% later decide whether we cite this or not 
%\footnote{The vector and scalar polarizations can exist even in GR when a GW propagates in a massive dispersive medium, though the effect is tiny \cite{Montani2019}.}
 while the scalar-tensor theory additionally introduces scalar degrees of freedom for a GW via scalar fields \cite{Eardley1973a, Will1993}. The scalar polarization modes of a GW are also predicted in $f(R)$ gravity \cite{Sotiriou2010}. It has been pointed out that scalar-tensor polarization mixing can occur in the gravitational lensing beyond GR \cite{Ezquiaga2020}.
%or beyond geometric optics \cite{Dalang:2021qhu}. 
 In this way, the degrees of freedom of a GW reflect the nature of gravity. Any signature of non-tensorial polarization modes that are forbidden in GR demands the extensions of GR if they are discovered \cite{Chen2021}.

The observations of GWs from compact binary coalescences \cite{TheLIGOScientificCollaboration2018b, Abbott2020b} by advanced LIGO \cite{Aasi2015} and advanced Virgo \cite{Acernese2015} have been able to test GR in a stronger gravity regime \cite{Abbott2016e,Abbott2020, Abbott2017b}. Some analytical and numerical studies to probe into the anomalous polarization modes have been made for 
%the different morphology of GWs: burst waves \cite{Hayama2013a}, stochastic backgrounds \cite{Nishizawa2009a,  Callister2017, Abbott2018b}, continuous waves \cite{Isi2015, Isi2017c},
GWs from compact binary coalescences \cite{Chatziioannou2012,Takeda2018,Takeda2019,Hagihara2019,Pang2020,Goyal:2020bkm,Chatziioannou2021}. So far, the polarization modes of GWs from compact binary coalescences have been tested in the pure polarization framework, in which one performs the model selection between GR and an extreme case of an alternative gravity theory allowing only scalar or vector polarization modes. In the context of such pure polarization tests, the signals of GW170814 \cite{Abbott2017b} and GW170817 \cite{TheLIGOScientificCollaboration2017b} strongly favor the tensor polarization against the pure vector or scalar polarizations \cite{Abbott2017b, TheLIGOScientificCollaboration2018c, Takeda2020}. On the other hand, the polarization test based on the null stream method that does not require any waveform of a GW has been performed \cite{Hagihara2019,Pang2020,Abbott2020}. The result supports GR modestly.

However, most of alternative theories of gravity predicts tensor modes along with subdominant vector and/or scalar modes. In other words, a detector signal would be typically a mixture of those polarization modes. As a waveform-free approach, the reconstruction method capable of detecting and characterizing the mixed polarizations of a transient GW signal has been proposed in \cite{Hayama2013a,Chatziioannou2021}. The most stringent constraint on the additional scalar amplitude has been obtained from the measurement of the orbital period of the binary pulsars and is consistent with the tensor modes in GR \cite{Will2005}. However, it is still in a weak gravity regime much before the merger of the binary. In this letter, we evaluate the amplitude of the scalar mode mixed with the ordinary tensor modes for GW170814 and GW170817, which are the events detected by three detectors so that we can separate a mixture of the scalar-tensor polarizations in principle~\cite{Takeda2018}. This is for the first time to add the scalar polarization term directly in the polarization test. Throughout this letter we use the geometrical units, $c=G=1$.

\section{Scalar-tensor polarization model}
\label{Scalar-tensor polarization model}

\subsection{Detector signal}
While GWs are allowed to have two tensorial polarization modes as the degrees of freedom of gravity in GR, a general metric theory of gravity allows at most four non-tensorial polarization modes in addition to tensorial modes ~\cite{Eardley1973a}. Thus, the detector signal of the I-th GW interferometric detector can be written in general as \cite{Will2005, Poisson2014}
 \begin{equation}
 \label{detector_signal}
 h_I(t,\hat{\Omega})=\sum_A F_I^{A}(\hat{\Omega})h_A(t),
 \end{equation} 
 where $h_{A}(t)$ are each polarization component of the GW, $\hat{\Omega}$ is the sky direction of a GW source, and $F_{I}^A$ is the antenna pattern functions of the I-th detector,
  \begin{equation}
  \label{antenna}
  F_I^{A}(\hat{\Omega}):=d_I^{ab}e^{A}_{ab}(\hat{\Omega}).
  \end{equation}
 Here, $A$ is polarization indices running over $+,\times, x, y, b, l$, corresponding to plus, cross, vector x, vector y, breathing, and longitudinal polarization modes, respectively. We can test gravity theories by separating polarizations of the observed signal \cite{Nishizawa2009a}. 

\subsection{Scalar-tensor waveform}
Since the two scalar polarizations (longitudinal and breathing) are completely degenerate, they can not be distinguished by the interferometric detectors \cite{Nishizawa2009a, Takeda2018}.
Thus, our scalar-tensor model should consist of three polarization modes, that is, plus, cross, and the dipole and quadrupole scalar polarization modes.

The angular pattern of GW emission, or the inclination-angle $\iota$ dependence of the scalar mode, should be proportional to $\sin{\iota}$ for the dipole radiation and $\sin^2{\iota}$ for the quadrupole radiation \cite{Chatziioannou2012, Takeda2018, Takeda2020}. In order to avoid introducing parameters that have different correlations with other parameters depending on the theory as free parameters, we assume the l-th harmonic in time domain $h^{(l)}(t)=\eta^{(2-l)/5}(\mathcal{M}/d_L)(2\pi\mathcal{M}F)^{l/3}e^{-il\Phi}$ with the inclination-dependence as the scalar-mode waveform. Here, $\mathcal{M}$ is the chirp mass, $d_L$ is the luminosity distance, and $\eta$ is the symmetric mass ratio. In addition, $F$ and $\Phi$ are the orbital frequency and phase, respectively. However, the tensor amplitude and the phasing should get a backreaction because the additional scalar radiation disrupts the orbital motion.

Therefore, we analyze the data under the scalar-tensor hypothesis $\HST$ having the following signal model in the frequency domain,
\begin{equation}
\begin{split}
 \tilde{h}_I(f,\hat{\Omega})&=\tilde{h}_{\text{GR}}(1+\delta A)e^{i\delta\Psi^{(2)}}\\
&+\sqrt{\frac{5\pi}{96}}A_{S2} F_{b}\sin^2{\iota}\frac{\mathcal{M}^2}{d_L}u_{2}^{-7/2}e^{-i\Psi_\text{GR}^{(2)}}e^{i\delta\Psi^{(2)}}\\
&+\sqrt{\frac{5\pi}{48}}A_{S1}F_{b}\sin{\iota}\eta^{1/5}\frac{\mathcal{M}^2}{d_L}u_{1}^{-9/2}e^{-i\Psi_\text{GR}^{(1)}}e^{i\delta\Psi^{(1)}},
\end{split}
\label{ST}
\end{equation}
where $\tilde{h}_{\rm GR}$ is the frequency-domain waveform of the GW from compact binary coalescences in GR and $\Psi_{\rm GR}^{(l)}$ is the frequency evolution for the $l$-th harmonic in GR. $u_l:=(2\pi\mathcal{M}f/l)^{1/3}$ is the the reduced $l$-th harmonic frequency.   %It is also assumed that the phase for the scalar mode is that of the plus mode. This is because the coalescence phase can hardly be determined even in GR, and the relative phase of the scalar mode is even more difficult to determine. 
$A_{S2}$ and $A_{S1}$ are the additional polarization parameters characterizing the quadrupole and dipole amplitudes of the scalar mode, respectively. $\delta A$ and $\delta \Psi$ denote the amplitude and phase corrections based on the energy loss due to the presence of the additional scalar radiation \cite{Chatziioannou2012}.  The additional scalar radiation requires the modification of the change rate of the binary binding energy as
\begin{equation}
\dot{E}=\dot{E}^{\text{GR}}\left(1+\frac{2}{3}A_{S2}^2+\frac{5}{96}A_{S1}^2\eta^{2/5}(2\pi\mathcal{M}F)^{-2/3}\right).
\end{equation}
where $\dot{E}^{(\rm{GR})}$ is the change rate in GR. Keeping up to the second order in terms of $A_{S}$, the modification leads to the amplitude and phase corrections as
\begin{equation}
\begin{split}
\delta A ^{(l)}&=\delta A_{q}^{(l)}+\delta A_{d}^{(l)}\\
&=-\frac{1}{3}A_{S2}^2-\frac{5}{192}A_{S1}^2\eta^{2/5}u_{l}^{-2},
\end{split}
\end{equation}
\begin{equation}
\begin{split}
\delta \Psi^{(l)}&=\delta \Psi_{q}^{(l)}+\delta \Psi_{d}^{(l)}\\
&=\frac{l}{128}A_{S2}^2 u_{l}^{-5}+\frac{5l}{14336}A_{S1}^2\eta^{2/5}u_{l}^{-7},
\end{split}
\end{equation}
through the stationary phase approximation. Consequently, we obtain the waveform model in \Eq{ST}.

 We deal with only the inspiral phase because the merger and ringdown waveforms in a scalar-tensor theory have not yet been constructed for use in a search due to the nonlinearity and complexity of the field equations, which might lead to large corrections in such stronger gravity regime \cite{Damour1996}. The dipole radiation has already well constrained by evaluating the -1PN order phase coefficient \cite{Abbott2020}. We can translate the constraint $|\delta \hat{\varphi}_{-2}|<10^{-5}$ for GW170817 \cite{TheLIGOScientificCollaboration2018c} into the dipole amplitude $A_{S1}\lesssim0.02$ assuming the relation between them in our model. Since this is sufficiently smaller than the amplitude determination accuracy expected from SNR, we ignore the dipole term for GW170817 considering the computational cost.

\section{Analysis}
\label{Analysis}
We analyze the signal of GW170814 \cite{Abbott2017b} and GW170817 \cite{Abbott2017} in a scalar-tensor polarization framework, which are GW events from a binary black-hole coalescence and a binary neutron-star coalescence observed by three-detector network, respectively. We basically take the same analysis method in \cite{Takeda2020} and analyze the data under the scalar-tensor hypothesis $\HST$, in which the GW signal is described as \Eq{ST}. In addition to $A_{S2}$ and $A_{S1}$, we consider 13 source parameters:
%\begin{equation}
%\bm{\theta}=(\alpha, \delta, \iota, \psi, d_L, \tc, \phic, m_1, m_2, \chi_1, \chi_2, \Lambda_1, \Lambda_2).
%\end{equation}
the right ascension and declination of the compact binary system, $\alpha$ and $\delta$, the inclination angle of the binary system $\iota$, the polarization angle $\psi$, the luminosity distance to the compact binary system $d_L$, the time and phase at the coalescence, $\tc$ and $\phic$,
detector-frame masses, $m_1$ and $m_2$, dimensionless spins of the primary and the secondary objects, $\chi_1$ and $\chi_2$, and the tidal deformability parameters of the primary and secondary stars, $\Lambda_1$ and $\Lambda_2$.

Our analysis relies on the Bayesian inference. The posterior probability distribution is calculated from the Bayes' theorem,
\begin{equation}
p(\bm{\theta}| \dIs, \HST) = \frac{p(\bm{\theta}) p(\dIs | \bm{\theta}, \HST)}{p(\dIs|\HST)}.
\end{equation}
$p(\bm{\theta})$ represents the prior probability distribution and is basically applied by the standard priors \cite{Abbott2019d} (see the dettail in \cite{, Takeda2020}).
Note that we apply a uniform priors in the range $[-1,1]$ for $A_{S2}$ and $A_{S1}$.
$p(\dIs | \bm{\theta}, \HST)$ represents the likelihood function and we apply the standard Gaussian noise likelihood \cite{Abbott2019d}.
The lower frequency cutoff for the likelihood calculations is $20\unit{Hz}$ for GW170814 and $23\unit{Hz}$ for GW170817 \cite{Abbott2019d}.

We utilize the Bilby software \cite{Ashton2019, Romero-Shaw2020} and the cpnest sampler \cite{Veitch2017} for the Bayesian inference.
As an inspiral template, we adopt TaylorF2 \cite{Buonanno2009}. For GW170817, we utilize the focused reduced order quadrature technique \cite{Morisaki2020} as in \cite{Takeda2020}. We use the data of GW170814 whose duration is 4 seconds and sampling frequency is $4096\unit{Hz}$ and the data of GW170817 with the removal of glitch whose duration is 128 seconds and sampling frequency is $4096\unit{Hz}$ from Gravitational Wave Open Science Center \cite{Abbott2019c}.

As for GW170817, an optical \cite{Coulter2017} and near-infrared \cite{Tanvir2017} electromagnetic counterpart was observed nearby the galaxy NGC 4993 ~\cite{Abbott2017c}, and the associated gamma ray burst, GRB 170817A, was also observed \cite{Meegan2009, Kienlin2003}. In the analysis of GW170817, we additionally impose the location priors on the luminosity distance, the right ascension, and the declination of GW170817 from the position of the host galaxy, NGC4993. The prior of the luminosity distance is set to be the Gaussian distribution with the mean 42.9 Mpc and the standard deviation 3.2 Mpc. We fix the right ascension and the declination to ${\rm RA} = 13{\rm h}09{\rm m}48{\rm s}.085$ and ${\rm DEC}=-23^{\circ}22'53''.343$ \cite{Abbott2017c}. In addition, the orientation of the compact binary system was also restricted for the viewing angle $\theta_{\rm obs}$ as $0.25\ {\rm rad}< \theta_{\rm obs} (d_L/41\ {\rm Mpc})< 0.45\ {\rm rad}$ \cite{Mooley2018, Hotokezaka2019}. This can be converted into the inclination angle, $\theta_{\rm obs}=\iota$ or $\theta_{\rm obs}=\pi-\iota$, from the assumption that the jet is perpendicular to the orbital plane of the compact binary system. Consequently, we impose the jet prior on the inclination angle in the range of $2.68\,{\rm rad} < \iota <2.92\, {\rm rad}$ \cite{Takeda2020}.

%%%%%%%%%%%%%%%%%%%%%%%%%%%%%%%%%%%%%%
\section{Result}
\label{Result}

%\subsubsection{General relativity}
%First, we analyze GW170814 in GR. In GR, the detector signal is expressed as
%\begin{equation}
 %h_I(t,\hat{\Omega})=F_I^{+}(\hat{\Omega})h_{+,{\rm GR}}(t)+F_I^{\times}(\hat{\Omega})h_{\times,{\rm GR}}(t).
%\end{equation}

%\begin{figure}
%\begin{center}
%\includegraphics[width=\hsize]{GW170814_tensor_TaylorF2.pdf}
%\end{center}
%\caption{GW170814, TaylorF2, GR}
%\label{GW170814_tensor_TaylorF2}
%\end{figure}

%\Fig{GW170814_tensor_TaylorF2} shows the results of the corner plots for the binary blackhole parameters with the TaylorF2 waveform in GR.
%A base $e$ logarithm of noise evidence is  -27767.896 and a logarithm of evidence is -27678.180 +/-  0.083. A logarithm of the Bayes factor is 89.716 +/-  0.083.

%\begin{figure}
%\begin{center}
%\includegraphics[width=\hsize]{tensor_scalar_GW170814_A_subset.pdf}
%\end{center}
%\caption{GW170814, TaylorF2, Scalar+Tensor}
%\label{tensor_scalar_GW170814_A_subset}
%\end{figure}

\begin{figure}
\begin{center}
\includegraphics[width=\hsize]{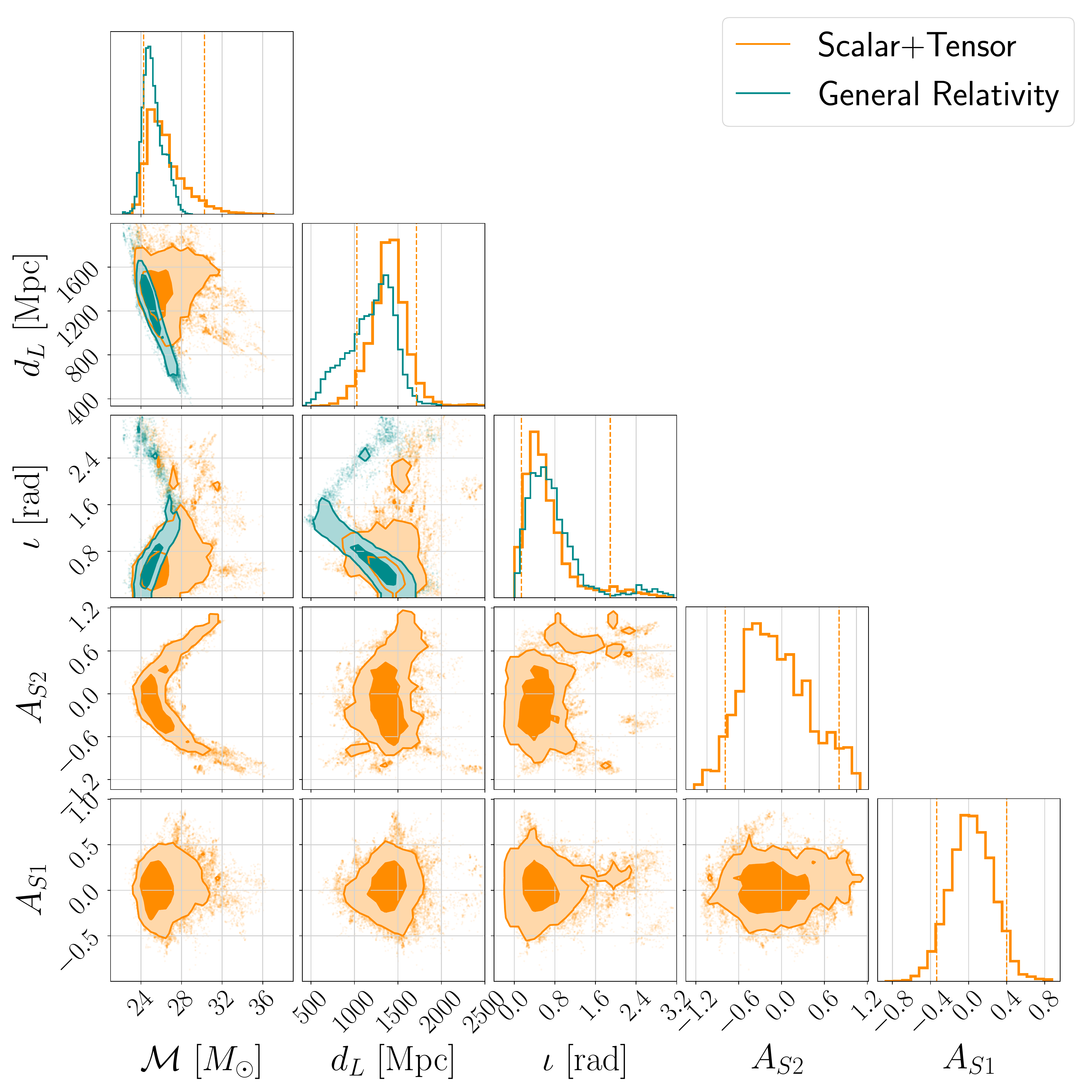}
\end{center}
\caption{The posterior probability distributions of the chirp mass in the source frame, the luminosity distance, the inclination angle, and the scalar amplitudes for GW170814 under the scalar-tensor hypothesis $\HST$. For comparison, the results under GR are also shown. The constraint on the scalar amplitude $A_{S2}$ can be converted into the constraint on the ratio of the scalar mode amplitude to the tensor mode amplitude: $R_{ST} \lesssim 0.20$.}
\label{GW170814_ST}
\end{figure}

%\begin{figure}
%\begin{center}
%\includegraphics[width=\hsize]{GW170814_GR.pdf}
%\end{center}
%\caption{The posterior probability distributions of the chirp mass in the source frame, the luminosity distance, and the inclination angle for GW170814 under GR.}
%\label{GW170814_GR}
%\end{figure}

\begin{figure}
\begin{center}
\includegraphics[width=\hsize]{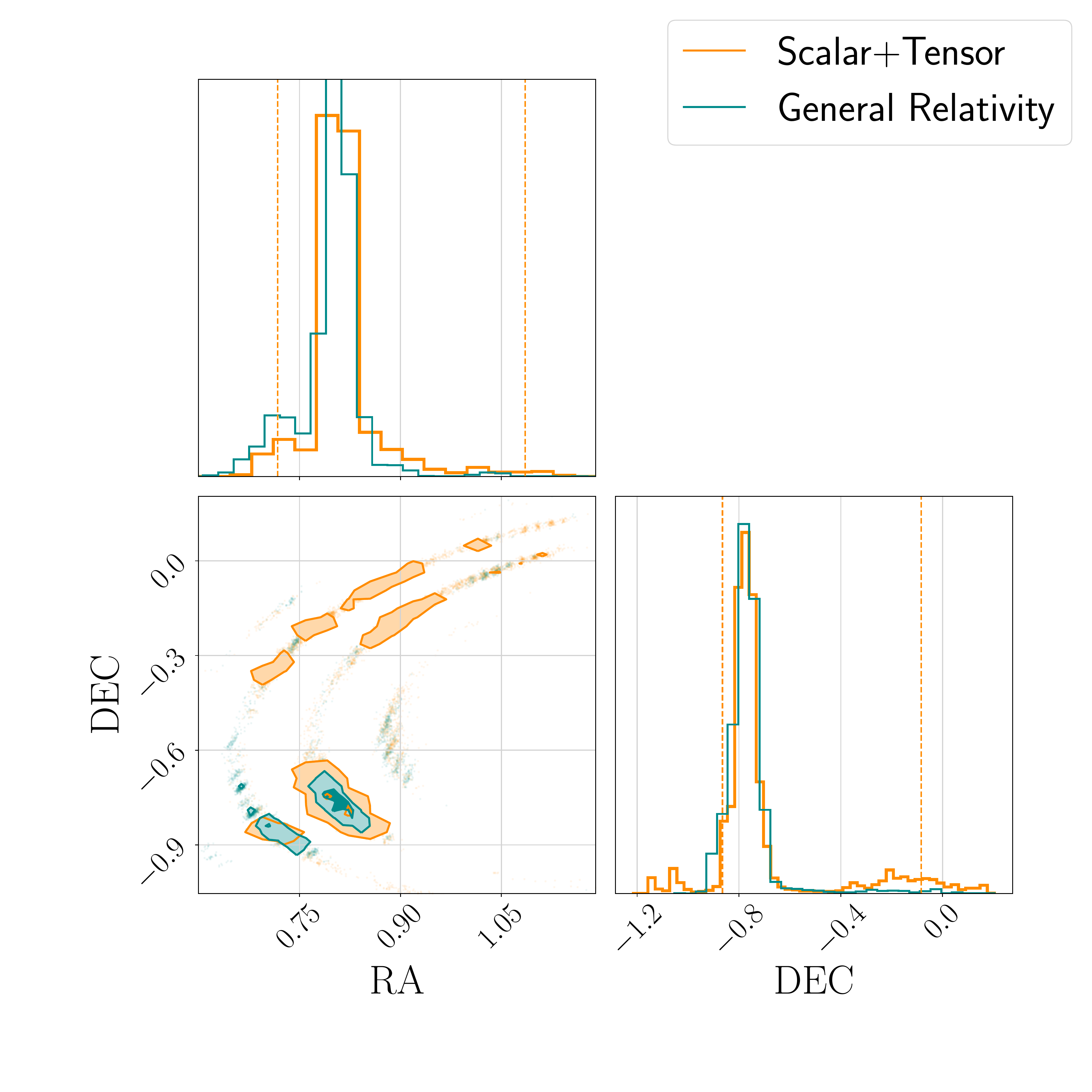}
\end{center}
\caption{The posterior probability distributions of GW170814 for the right ascension and the declination are shown in orange under the scalar-tensor hypothesis $\HST$ and in green under GR.}
\label{GW170814_subset_skymap_paper}
\end{figure}

%botsu
%\begin{figure}
%\begin{center}
%\includegraphics[width=\hsize]{tensor_scalar_GW170814_A_phase.pdf}
%\end{center}
%\caption{Tensor + scalar on GW170814 with A and phase for scalar mode.}
%\label{tensor_scalar_GW170814_A_phase}
%\end{figure}

%\Fig{tensor_scalar_GW170814_A_subset} shows the result.
For GW170814, \Fig{GW170814_ST} shows the posterior probability distribution for the chirp mass in the source frame, the luminosity distance, the inclination angle, and the scalar amplitudes under $\HST$. \Fig{GW170814_subset_skymap_paper} shows the posterior probability distribution of the right ascension and the declination under $\HST$ in orange. The results under GR are also shown in green for comparison. These amplitude parameters and the sky location are almost unchanged by adding the scalar mode. The amplitude parameters for the additional scalar modes are constrained by $-0.13^{+0.89}_{-0.63}$ for the quadrupole and $0.03^{+0.37}_{-0.37}$ for the dipole. From the comparison of the estimated errors from the scalar amplitude in the scalar term and the phase correction of the tensor modes depending on the scalar amplitude, we can see that the contribution from the scalar term dominates compared to the contribution from the phase modification for the tensor mode when the amplitude is small $A_{S2}\lesssim0.5$. \Fig{GW170814_ST} shows that the scalar term starts to be constrained in such a region of small scalar amplitude, shifting the posterior distribution to the negative side slightly. This is because the scalar term compensates for the change in amplitude of the tensor mode with increasing mass, as shown by the 2D correlations. We can also see the 2D correlations between $A_{S2}$ and ${\cal M}$ or $d_L$. The correlations are interpreted from the parameter combination, $(1-{A_{S2}}^2/3)\mathcal{M}^{5/6}/d_L$, in the tensor-mode amplitude, which dominates the scalar-mode amplitude. This combination of parameters is tightly constrained by the observational data, thereby introducing correlations between the individual parameters.

\begin{figure}
\begin{center}
\includegraphics[width=\hsize]{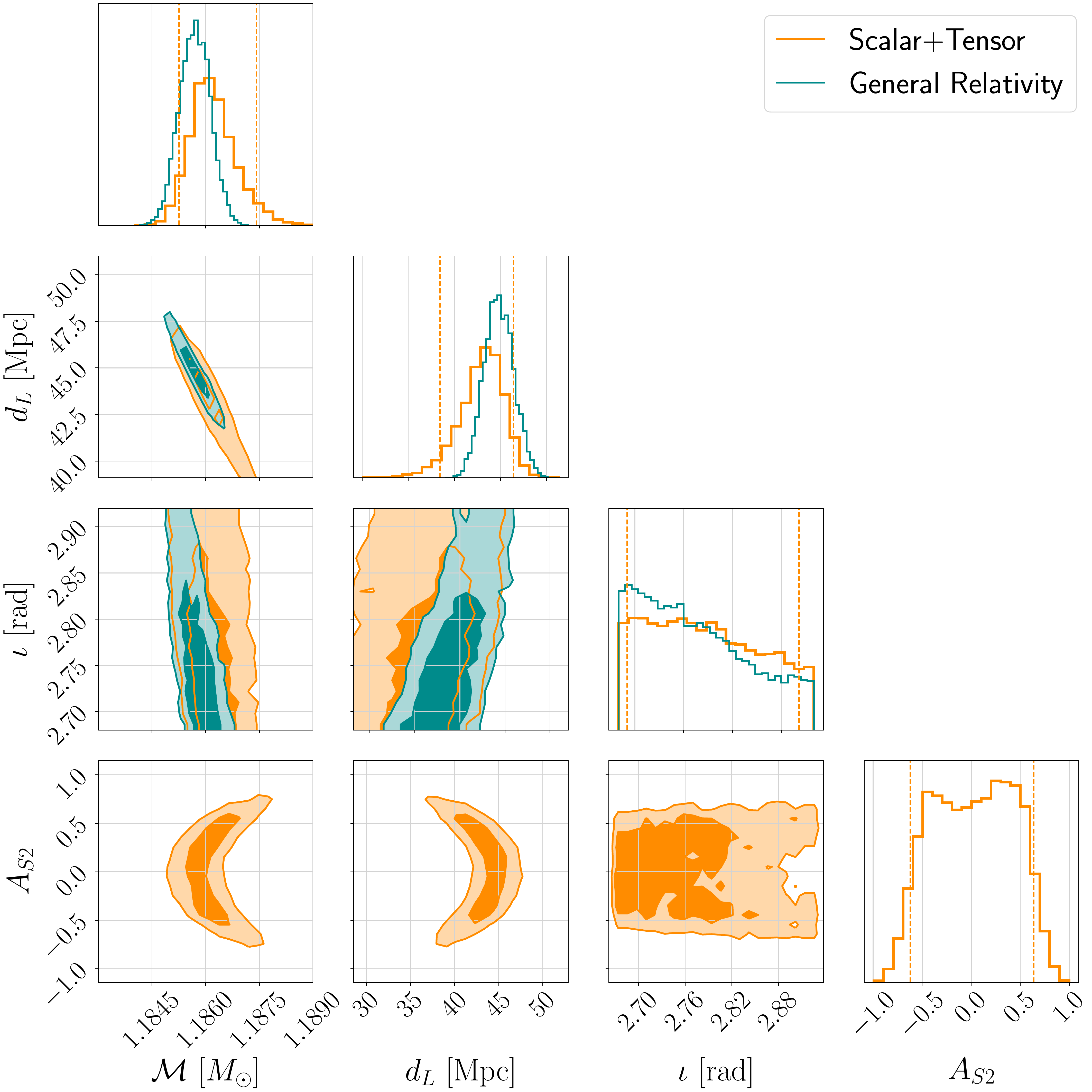}
\end{center}
\caption{The posterior probability distributions of the chirp mass in the source frame, the luminosity distance, the inclination angle, and the scalar amplitude for GW170817 under the scalar-tensor hypothesis $\HST$. For comparison, the results under GR are also shown. The constraint on the scalar amplitude $A_S$ can be converted into the constraint on the ratio of the scalar mode amplitude to the tensor mode amplitude: $R_{ST} \lesssim 0.068$.}
\label{GW170817_ST}
\end{figure}

For GW170817, \Fig{GW170817_ST} shows the posterior probability distributions for the chirp mass in the source frame, the luminosity distance, the inclination angle, and the scalar amplitude. The distributions of the amplitude parameters are hardly different between under GR and under $\HST$. For GW170817, the stronger correlations between $A_{S2}$ and ${\cal M}$ or $d_L$ are seen. The amplitude parameter for an additional scalar mode is constrained in $0.04^{+0.60}_{-0.66}$.

The constraints can be translated into the constraints on the ratio of the scalar amplitude to the tensor amplitude defined for our scalar-tensor waveform model by \cite{Yang2017b}
\begin{equation}
R_{ST}:=\left|\frac{A_{\rm S2} \sin^2 \iota}{\sqrt{(1+\cos^2 \iota)^2/4+\cos^2 \iota}}\right|.
\end{equation}
This ratio represents how deep the scalar mode is searched in a GW signal.
We find the constraints on $R_{\text ST}$:
$R_{\rm ST} \lesssim 0.20$
%\begin{equation}
%R_{\rm ST}=-0.02^{+0.17}_{-0.18},
%\end{equation}
for GW170814 and 
$R_{\rm ST} \lesssim 0.068$
%\begin{equation}
%R_{\rm ST}=0.004^{+0.062}_{-0.069},
%\end{equation}
for GW170817, which are consistent with GR. This is thought to be because if more scalar waves are emitted, the tensor mode is also significantly deformed and the observed signal cannot be explained. Concomitantly, the amplitude parameters and the location parameters are hardly changed. We confirmed that the results do not change significantly with increased number of live points. The precision of the additional amplitude for GW170814 and GW170817 is comparable, while the amplitude ratio is better determined for GW170817. This comes from the fact that the inclination angle estimated from the gamma ray burst is nearly face-on and the factor $\sin^2 \iota$ in the numerator becomes small. The constraint on the scalar coupling in the alternative theories of gravity is given by the additional scalar amplitude $A_{S2}$, while the amplitude ratio $R_{ST}$ can be regarded as the indicator of the search depth to the polarization modes because the smaller $R_{ST}$ is, the deeper we are able to explore the scalar mode, given the inclination angle.

%We obtain logarithm of the Bayes factor between $\HST$ and the noise hypothesis of $93.4$, while we find logarithm of the Bayes factor between GR and the noise hypothesis of $89.7$.

%A base $e$ logarithm of noise evidence is  -27767.896 and a logarithm of evidence is -27674.508 +/-  0.080. Thus, a logarithm of the Bayes factor is 93.388 +/-  0.0803.The amplitude parameter for additional scalar mode is constrained by $3.44^{+3.64}_{-5.06}$.  

%\subsubsection{General relativity}
%First, we analyze GW170817 in GR. \Fig{GW170817_tensor_TaylorF2} shows the results of the corner plots for the binary blackhole parameters with the TaylorF2 waveform in GR. A base e logarithm of noise evidence is -724566.347 and a logarithm of evidence is -724057.339. A logarithm of the Bayes factor is 509.009.

%\begin{figure}
%\begin{center}
%\includegraphics[width=\hsize]{GW170817_tensor_TaylorF2.pdf}
%\end{center}
%\caption{GW170817, TaylorF2, GR}
%\label{GW170817_tensor_TaylorF2}
%\end{figure}

%\subsubsection{Scalar-tensor}

%\begin{figure}
%\begin{center}
%\includegraphics[width=\hsize]{tensor_scalar_GW170817_A_subset.pdf}
%\end{center}
%\caption{GW170817, TaylorF2, Scalar+Tensor}
%\label{tensor_scalar_GW170817_A_subset}
%\end{figure}

%\begin{figure}
%\begin{center}
%\includegraphics[width=\hsize]{GW170817_GR.pdf}
%\end{center}
%\caption{The posterior probability distributions of the chirp mass in the source frame, the luminosity distance, and the inclination angle for GW170817 under GR.}
%\label{GW170817_GR}
%\end{figure}

%\Fig{tensor_scalar_GW170817_A_subset} shows the result.
 %base $e$ logarithm of noise evidence is  -724566.347 and a logarithm of evidence is -724057.918. Thus, a logarithm of the Bayes factor is 508.430.The amplitude parameter for additional scalar mode is constrained by $0.58^{+7.61}_{-8.70}$.  \Fig{GW170817_subset_chirp_paper} shows the comparison between the results of the tensor framework and the scalar-tensor framework. 

 %This is likely due to the fact that imposing the location prior and the jet prior reduces the number of parameters that can compensate for each other. 

 %We obtain logarithm of the Bayes factor between $\HST$ and the noise hypothesis of $508.43$, while we find logarithm of the Bayes factor between GR and the noise hypothesis of $509.01$.

\section{Conclusion and Discussions}
\label{CD}

We searched for a mixed scalar-tensor polarization modes of GW170814 and GW170817. We found the constraints on the additional scalar polarization amplitudes, $A_{S2}=-0.13^{+0.89}_{-0.63}$ and $A_{S1}=0.03^{+0.37}_{-0.37}$ for GW170814 and $A_{S2}=0.04^{+0.60}_{-0.66}$ for GW170817. These results can be translated into the constraints on the ratio of the scalar amplitude to the tensor amplitude in the GW signals: $R_{\rm ST} \lesssim 0.20$ for GW170814 and $R_{\rm ST} \lesssim 0.068$ for GW170817. Note that since the radiation mechanisms of different physical systems such as the binary black holes and the binary neutron stars are not necessarily the same, the two constraints could not be simply compared. 

When the fourth and fifth detector such as KAGRA \cite{Somiya2012, KAGRACollaboration2020, KAGRACollaboration2013} and LIGO India \cite{Iyer2011} participates in the GW detector network, four and five polarization modes can be probed \cite{Takeda2018}. Therefore, the expansion of the detector network will make it possible to probe the anomalous GW polarizations directly under a vector-tensor framework or a scalar-vector-tensor framework in the future. On the other hand, the detection limit for the additional amplitude is determined by the SNR. Thus, it is expected that the event with the large SNR can be utilized for more precise tests. However, since there is a limit to the accuracy of test by a single GW, we are developing a method to statistically combine the polarization analysis results of multiple GW sources.

\section*{Acknowledgements}
H. T. acknowledges financial support received from the Advanced Leading Graduate Course for Photon Science (ALPS) program at the University of Tokyo. H.T. is supported by JSPS KAKENHI Grant No. 18J21016. S. M. is supported by JSPS KAKENHI Grant No. 19J13840 and NSF PHY-1912649. A.~N. is supported by JSPS KAKENHI Grant Nos. JP19H01894 and JP20H04726 and by Research Grants from Inamori Foundation.This research has made use of data, software and/or web tools obtained from the Gravitational Wave Open Science Center (https://www.gw-openscience.org/ ), a service of LIGO Laboratory, the LIGO Scientific Collaboration and the Virgo Collaboration. LIGO Laboratory and Advanced LIGO are funded by the United States National Science Foundation (NSF) as well as the Science and Technology Facilities Council (STFC) of the United Kingdom, the Max-Planck-Society (MPS), and the State of Niedersachsen/Germany for support of the construction of Advanced LIGO and construction and operation of the GEO600 detector. Additional support for Advanced LIGO was provided by the Australian Research Council. Virgo is funded, through the European Gravitational Observatory (EGO), by the French Centre National de Recherche Scientifique (CNRS), the Italian Istituto Nazionale di Fisica Nucleare (INFN) and the Dutch Nikhef, with contributions by institutions from Belgium, Germany, Greece, Hungary, Ireland, Japan, Monaco, Poland, Portugal, Spain. \newpage


\bibliographystyle{h-physrev3}

\begin{thebibliography}{10}

\bibitem{Will1993}
C.~M. Will,
\newblock {\em {Theory and Experiment in Gravitational Physics}} (Cambridge
  University Press, 1993).

\bibitem{Stairs2003}
I.~H. Stairs,
\newblock Living Reviews in Relativity {\bf 6}, 5 (2003).

\bibitem{Murata2015}
J.~Murata and S.~Tanaka,
\newblock Classical and Quantum Gravity {\bf 32}, 033001 (2015).

\bibitem{Will2005}
C.~M. Will,
\newblock Living Reviews in Relativity {\bf 9}, 3 (2006), 0510072.

\bibitem{Clifton2012}
T.~Clifton, P.~G. Ferreira, A.~Padilla, and C.~Skordis,
\newblock Physics Reports {\bf 513}, 1 (2012),
\newblock Modified Gravity and Cosmology.

\bibitem{Brans1961}
C.~Brans and R.~H. Dicke,
\newblock Physical Review {\bf 124}, 925 (1961).

\bibitem{Fujii2003}
Y.~Fujii and K.-i. Maeda,
\newblock Classical and Quantum Gravity {\bf 20}, 4503 (2003).

\bibitem{Perrotta1999}
F.~Perrotta, C.~Baccigalupi, and S.~Matarrese,
\newblock Phys. Rev. D {\bf 61}, 023507 (1999).

\bibitem{Eardley1973a}
D.~M. Eardley, D.~L. Lee, A.~P. Lightman, R.~V. Wagoner, and C.~M. Will,
\newblock Physical Review Letters {\bf 30}, 884 (1973).

\bibitem{Sotiriou2010}
T.~P. Sotiriou and V.~Faraoni,
\newblock Reviews of Modern Physics {\bf 82}, 451 (2010).

\bibitem{Ezquiaga2020}
J.~M.~Ezquiaga, M.~Zumalac\'{a}rregui,
\newblock Physical Review D {\bf 102}, 124048 (2020).

%\bibitem{Dalang:2021qhu}
%C.~Dalang, G.~Cusin, and M.~Lagos,
%arXiv:2104.10119 (2021).

%\bibitem{Montani2019}
%G.~Montani and F.~Moretti,
%\newblock Physical Review D {\bf 100}, 024045 %(2019).

%\bibitem{Callister2017}
%T.~Callister {\em et~al.},
%\newblock Physical Review X {\bf 7}, 041058 (2017).

%\bibitem{Abbott2018b}
%B.~Abbott {\em et~al.},
%\newblock Physical Review Letters {\bf 120}, 201102 (2018).

\bibitem{Chen2021}
Z.C.~Chen, C.~Yuan, and Q.G.~Huang,
\newblock arXiv:2101.06869 (2021).

%\bibitem{Isi2015}
%M.~Isi, A.~J. Weinstein, C.~Mead, and M.~Pitkin,
%\newblock Physical Review D {\bf 91}, 082002 (2015).

%\bibitem{Isi2017c}
%M.~Isi, M.~Pitkin, and A.~J. Weinstein,
%\newblock Physical Review D {\bf 96}, 042001 (2017), 1703.07530.

\bibitem{TheLIGOScientificCollaboration2018b}
{The LIGO Scientific Collaboration} {\em et~al.},
\newblock Physical Review X {\bf 9}, 031040 (2018), 1811.12907.

\bibitem{Abbott2020b}
R.~Abbott {\em et~al.},
\newblock arXiv:2010.14527 (2021).

\bibitem{Aasi2015}
J.~Aasi {\em et~al.},
\newblock Classical and Quantum Gravity {\bf 32}, 074001 (2015).

\bibitem{Acernese2015}
F.~Acernese {\em et~al.},
\newblock Classical and Quantum Gravity {\bf 32}, 024001 (2015).

\bibitem{Abbott2016e}
B.~P. Abbott {\em et~al.},
\newblock Physical Review X {\bf 6}, 041015 (2016).

\bibitem{Abbott2020}
LIGO Scientific Collaboration and Virgo Collaboration, R.~Abbott {\em et~al.},
\newblock arXiv:2010.14529 (2020).

\bibitem{Abbott2017b}
B.~P. Abbott {\em et~al.},
\newblock Physical Review Letters {\bf 119}, 141101 (2017).


\bibitem{Chatziioannou2012}
K.~Chatziioannou, N.~Yunes, and N.~Cornish,
\newblock Physical Review D {\bf 86}, 022004 (2012).

\bibitem{Takeda2018}
H.~Takeda {\em et~al.},
\newblock Physical Review D {\bf 98}, 022008 (2018), 1806.02182.

\bibitem{Takeda2019}
H.~Takeda {\em et~al.},
\newblock Physical Review D {\bf 100}, 042001 (2019).

\bibitem{Hagihara2019}
Y.~Hagihara, N.~Era, D.~Iikawa, A.~Nishizawa, and H.~Asada,
\newblock Physical Review D {\bf 100}, 064010 (2019).

\bibitem{Pang2020}
P.~T.~H. Pang, R.~K.~L. Lo, I.~C.~F. Wong, T.~G.~F. Li, and C.~Van Den~Broeck,
\newblock Phys. Rev. D {\bf 101}, 104055 (2020).

\bibitem{Goyal:2020bkm}
S.~Goyal, K.~Haris, A.~K.~Mehta, P.~Ajith
\newblock Phys. Rev. D {\bf 103}, 024038 (2021).

\bibitem{Chatziioannou2021}
K.~Chatziioannou, M.~Isi, C.J.~Haster, and T.B. Littenberg
\newblock arXiv:2105.01521 (2021).


\bibitem{TheLIGOScientificCollaboration2017b}
LIGO Scientific Collaboration and Virgo Collaboration, B.~P. Abbott {\em
  et~al.},
\newblock Phys. Rev. Lett. {\bf 119}, 161101 (2017).

\bibitem{TheLIGOScientificCollaboration2018c}
{The LIGO Scientific Collaboration} {\em et~al.},
\newblock Physical Review Letters {\bf 123}, 011102 (2018), 1811.00364.

\bibitem{Takeda2020}
H.~Takeda, S.~Morisaki, and A.~Nishizawa,
%\newblock Pure polarization test of gw170814 and gw170817 using waveforms consistent with modified theories of gravity, 2020, 2010.14538.
\newblock Phys. Rev. D {\bf 103}, 064037 (2021).

\bibitem{Hayama2013a}
K.~Hayama and A.~Nishizawa,
\newblock Physical Review D {\bf 87}, 062003 (2013).

\bibitem{Poisson2014}
E.~Poisson and C.~M. Will,
\newblock {\em {Gravity}} (Cambridge University Press, Cambridge, 2014).

\bibitem{Nishizawa2009a}
A.~Nishizawa, A.~Taruya, K.~Hayama, S.~Kawamura, and M.-a. Sakagami,
\newblock Physical Review D {\bf 79}, 082002 (2009).

\bibitem{Damour1996}
T.~Damour and G.~Esposito-Far\`ese,
\newblock Phys. Rev. D {\bf 54}, 1474 (1996).

\bibitem{Abbott2017}
B.~P. Abbott {\em et~al.},
\newblock Physical Review Letters {\bf 119}, 161101 (2017).

\bibitem{Abbott2019d}
LIGO Scientific Collaboration and Virgo Collaboration, B.~P. Abbott {\em
  et~al.},
\newblock Phys. Rev. X {\bf 9}, 031040 (2019).

\bibitem{Ashton2019}
G.~Ashton {\em et~al.},
\newblock The Astrophysical Journal Supplement Series {\bf 241}, 27 (2019).

\bibitem{Romero-Shaw2020}
I.~M. Romero-Shaw {\em et~al.},
\newblock Monthly Notices of the Royal Astronomical Society {\bf 499}, 3295
  (2020),
  https://academic.oup.com/mnras/article-pdf/499/3/3295/34052625/staa2850.pdf.

\bibitem{Veitch2017}
J.~Veitch, W.~D. Pozzo, and C.~M. Pitkin,
\newblock https://doi.org/10.5281/ zenodo.825456  (2017).

\bibitem{Buonanno2009}
A.~Buonanno, B.~R. Iyer, E.~Ochsner, Y.~Pan, and B.~S. Sathyaprakash,
\newblock Phys. Rev. D {\bf 80}, 084043 (2009).

\bibitem{Morisaki2020}
S.~Morisaki and V.~Raymond,
\newblock Phys. Rev. D {\bf 102}, 104020 (2020).

\bibitem{Abbott2019c}
LIGO Scientific, Virgo, R.~Abbott {\em et~al.},
\newblock SoftwareX {\bf 13}, 100658 (2021), 1912.11716.

\bibitem{Coulter2017}
D.~A. Coulter {\em et~al.},
\newblock Science {\bf 358}, 1556 (2017).

\bibitem{Tanvir2017}
N.~R. Tanvir {\em et~al.},
\newblock The Astrophysical Journal {\bf 848}, L27 (2017).

\bibitem{Abbott2017c}
B.~P. Abbott {\em et~al.},
\newblock The Astrophysical Journal {\bf 848}, L12 (2017).

\bibitem{Meegan2009}
C.~Meegan {\em et~al.},
\newblock The Astrophysical Journal {\bf 702}, 791 (2009).

\bibitem{Kienlin2003}
A.~von Kienlin {\em et~al.},
\newblock Astronomy \& Astrophysics {\bf 411}, L299 (2003).

\bibitem{Mooley2018}
K.~P. Mooley {\em et~al.},
\newblock Nature {\bf 561}, 355 (2018).

\bibitem{Hotokezaka2019}
K.~Hotokezaka {\em et~al.},
\newblock Nature Astronomy {\bf 3}, 940 (2019).

\bibitem{Yang2017b}
H.~Yang, A.~Nishizawa, and U.-L. Pen,
\newblock Phys. Rev. D {\bf 95}, 084049 (2017).

\bibitem{Somiya2012}
K.~Somiya,
\newblock Classical and Quantum Gravity {\bf 29}, 124007 (2012).

\bibitem{KAGRACollaboration2020}
{KAGRA Collaboration} {\em et~al.},
\newblock (2020), 2005.05574.

\bibitem{KAGRACollaboration2013}
{KAGRA Collaboration} {\em et~al.},
\newblock Physical Review D {\bf 88}, 043007 (2013).

\bibitem{Iyer2011}
B.~Iyer {\em et~al.},
\newblock LIGO-India Tech. rep.
  (2011),
  https://dcc.ligo.org/LIGO-M1100296/public.

\end{thebibliography}

\end{document}